\documentclass[aip,apl,amsmath,floatfix,reprint]{revtex4-1}
\usepackage{graphicx,psfrag,mathptmx}

\begin{document}

\title{Reduced leakage current in Josephson tunnel junctions with codeposited barriers}
\author{Paul B. Welander}
    \altaffiliation{Present address: Lincoln Laboratory, Massachusetts Institute of Technology, Lexington, MA 02420}
\author{Timothy J. McArdle}
    \altaffiliation{Present address: IBM Thomas J. Watson Research Center, Yorktown Heights, NY 10598}
\author{James N. Eckstein}\email{eckstein@illinois.edu}
\affiliation{Department of Physics and Frederick Seitz Materials Research Laboratory,
             University of Illinois at Urbana-Champaign, Urbana, IL 61801}
\date{December 15, 2010}
\begin{abstract}
Josephson junctions were fabricated using two different methods of barrier formation.  The trilayers employed were Nb/Al-AlO$_x$/Nb on sapphire, where the first two layers were epitaxial. The oxide barrier was formed either by exposing the Al surface to O$_2$ or by codepositing Al in an O$_2$ background. The codeposition process yielded tunnel junctions that showed the theoretically predicted subgap current and no measurable shunt conductance. In contrast, devices with barriers formed by thermal oxidation showed a small shunt conductance in addition to the predicted subgap current.
\end{abstract}
\maketitle

Josephson tunnel junctions work by connecting the collective superconducting order of one electrode with that of another via elastic tunneling through a weak link.\cite{Josephson1962,Likharev1979}  The most robust process devised to date uses the diffusion-limited oxidation of an Al base layer to form a relatively uniform amorphous oxide on top of which a second superconducting film can be deposited.\cite{Gurvitch1983}  For operation at higher temperatures, Nb can be used for the electrode layers with only a thin Al layer for oxide barrier formation.  As components in macroscopic quantum circuits, however, these junctions do not perform as well as desired.  Spurious two-level systems arising from materials defects are viewed as the primary cause,\cite{Simmonds2004} but it seems clear that non-coherent resistive shunting of the junction is also a source of decoherence.  Buhrman and co-workers have studied the diffusion process and concluded that oxygen vacancies caused by inhomogeneous oxidation can lead to a measurable density of mid-gap states.\cite{Tan2005}  The density of these vacancies can be substantially reduced by several treatments that effectively activate diffusion kinetics.\cite{Mather2005}

Another way to improve the kinetics of the oxidation reaction is to \textit{grow} the oxide layer-by-layer.  In this Letter we show that resistive transport through a shunting channel can be eliminated by growing the amorphous AlO$_x$ layer-by-layer at room temperature by codeposition of Al and O$_2$.  The current-voltage ($I$-$V$) characteristics obtained from such junctions conform to predictions for ideal tunneling behavior in which the subgap current is due to thermally occupied quasiparticle states in the superconducting electrodes.\cite{VanDuzer1998}  By eliminating incoherent coupling modes between the electrodes, longer coherence times may be achieved and such junctions may be useful as components in quantum circuits.

Tunnel junctions were fabricated from trilayer samples grown in a molecular beam epitaxy (MBE) system that has been described previously.\cite{Welander2007}  Each trilayer had a single-crystal base electrode consisting of a 1000~{\AA}-thick Nb (110) film, grown at 800~$^{\circ}$C on A-plane sapphire, and a heteroepitaxial Al (111) layer grown at room temperature.  Al was deposited at a rate of 1-3~{\AA}/s with the layer thickness varied from 60-520~{\AA}.  Formation of an oxide tunnel barrier, as described below, was followed by the room temperature deposition of a poly-crystalline Nb top electrode.  Subsequent device fabrication utilized a selective Nb etch process with self-aligned SiO$_2$ dielectric and a Nb wiring layer.\cite{WelanderThesis}

The oxide layer was manufactured in one of two ways, both at room temperature. The first involved exposing the Al (111) surface to O$_2$ in a connected chamber for one hour at 10 torr.  This is the conventional approach developed by Gurvitch \textit{et al.},\cite{Gurvitch1983} where the thickness of the oxide is determined by a combination of exposure and diffusion kinetics.  The second process involved the codeposition of Al and O$_2$ at room temperature on the Al (111) surface in the MBE chamber.\cite{Welander2007}  The O$_2$ partial pressure during codeposition was $5\times10^{-6}$ torr.  In this case, the stoichiometry of the as-grown oxide was limited by the oxidation equilibrium on the growing surface.  These samples were further exposed to O$_2$ as described above.

The trilayer samples were studied using \textit{in situ} reflection high-energy electron diffraction (RHEED) and \textit{ex situ} atomic force microscopy (AFM) and x-ray diffraction (XRD).  Diffraction measurements confirmed the epitaxial nature of the Nb/Al bi-layers.  Al (111) was found to grow on Nb (110) in the Nishiyama-Wassermann orientation,\cite{Bruce1978} with Al $[1\bar{1}1]$ $\parallel$ Nb [001], consistent with previously published reports.\cite{Tao1984,Yamamoto2002}  XRD analysis also revealed twinning in the Al film arising from the two fcc (111) stacking variants.

\begin{figure}[t]
  \includegraphics[width=3.25in]{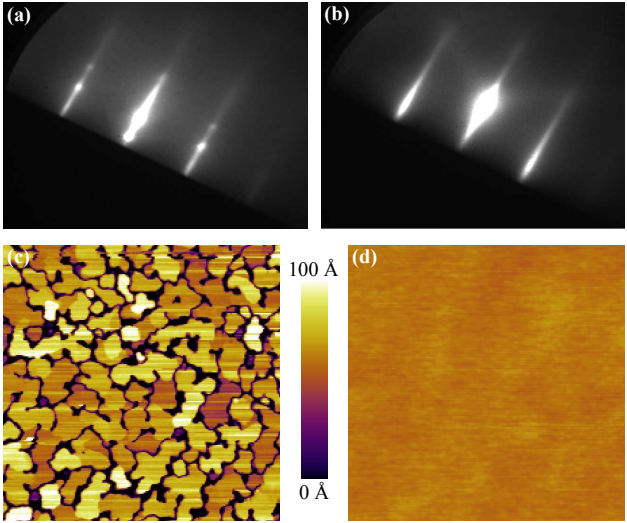}
  \caption{(Color online) RHEED (a) and AFM (c) from a 65~{\AA} Al (111) film grown on Nb (110).  Panels (b) and (d) show similar measurements made on a 200~{\AA} thick Al layer.  AFM scan areas are 2 $\times$ 2 $\mu$m$^2$.}\label{alcomp}
\end{figure}

RHEED patterns showed an evolution of the Al surface morphology during deposition.  Diffraction images from the Al (111) surface are shown in Fig. \ref{alcomp}.  Thinner films (less than about 100~{\AA}) showed superimposed RHEED and transmission patterns, indicating a mixed morphology of flat regions and islands.  With increasing thickness the transmission spots disappeared leaving only the RHEED streaks indicative of a flat surface.  RHEED measurements also indicated that the AlO$_x$ layers were amorphous, regardless of the process details.

AFM measurements confirmed the morphological evolution of the epitaxial Al films.  Figure \ref{alcomp} shows the surface morphology of Al (111) layers measuring 65~{\AA} and 200~{\AA} in thickness.  Thicker Al layers typically had an rms roughness of less than 5~{\AA}, while thinner ones commonly showed pin holes that penetrated the entire film thickness. (Because of the similar growth conditions and film thickness, we suspect that it was this lack of Nb coverage that may have caused the devices of Braginski \textit{et al.}\cite{Braginski1986} to exhibit leakage.)

\begin{figure}[b]
  \includegraphics[width=3.25in]{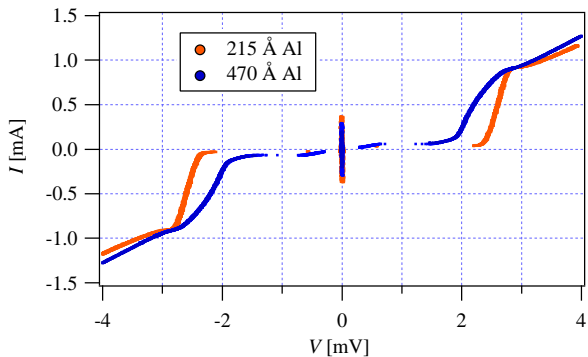}
  \caption{(Color online) Tunnel junction $I$-$V$ characteristics measured at 4.2~K for $10\times10$ $\mu$m$^2$ devices with differing Al layer thicknesses.  Both had diffused oxide barriers, and the only substantial difference was the fall-off width in the quasiparticle curve at the gap voltage.}\label{althickvar}
\end{figure}

Because superconductivity in the Al layer relies on the proximity effect at 4.2~K, devices with different Al thicknesses should exhibit contrasting behavior.  In Fig. \ref{althickvar} the $I$-$V$ characteristics of two $10\times10$ $\mu$m$^2$ devices measured at 4.2~K are shown, one with 215~{\AA} of Al and the other with 470~{\AA}.  The most striking difference is the voltage range over which quasiparticle tunneling turns off at the gap voltage.  A narrower turn-off was observed for the device with 215~{\AA} of Al -- about 0.5~mV -- compared with approximately 1.0~mV for the thicker Al layer.  Outside of the subgap structure there was little that distinguished these two samples.  Both curves showed a ``knee" in the quasiparticle curve at 2.8 mV, common for Nb-based Josephson junctions.  The critical currents, $I_c$, and normal-state resistances, $R_n$, were also comparable: 0.37~mA and 3.4~$\Omega$ (215~{\AA} Al) versus 0.29~mA and 2.9~$\Omega$ (470~{\AA} Al).

For comparing diffused and codeposited oxide barriers, three trilayer samples were grown, all having 200~{\AA} of Al but with different barriers: (1) thermal oxidation of the Al surface; (2) 15~{\AA} of codeposited AlO$_x$ plus thermal oxidation; (3) 20~{\AA} of codeposited AlO$_x$ plus thermal oxidation.  Tunnel junctions fabricated from these trilayers exhibited a range of critical current densities from 350 A/cm$^2$ for process (1), to 160 and 25 A/cm$^2$ for processes (2) and (3), respectively.

\begin{figure}[t]
  \includegraphics[width=3.25in]{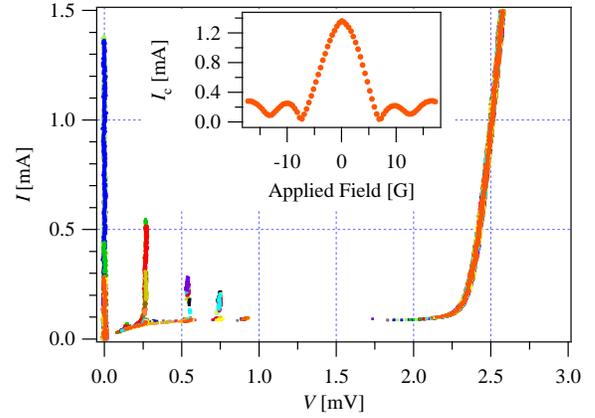}
  \caption{(Color online) Fiske modes for a $30\times30$ $\mu$m$^2$ device with 15~{\AA} of codeposited AlO$_x$.  Colors represent $I$-$V$s taken at 4.2~K with different applied fields in the range 0-17~G.  The reentrant shape of the subgap current leads to retrapping even when the critical current is nulled.  The inset shows the modulation of $I_c$ with field strength.}\label{fiske}
\end{figure}

These junctions were subjected to transverse magnetic fields, and the field-dependence of $I_c$ took the familiar Fraunhofer form.  Fiske modes, indicative of highly uniform tunnel barriers, were also observed for the devices with codeposited oxides.\cite{Fiske1964,Eck1964,Kulik1965}  The $I$-$V$ traces shown in Fig. \ref{fiske} are from a $30\times30$ $\mu$m$^2$ device with a 15~{\AA} codeposited barrier, and demonstrate these resonances for applied fields ranging from 0-17~G.  Over this range, Fiske modes out to at least 3rd order were excited.

The voltages, $V_n$, at which Fiske modes occur are related to the superconducting penetration depth, $\lambda$, by the expression
\[ V_n = \frac{nhv_{\textrm{ph}}}{4eL} = \frac{nhc}{4eL}\sqrt{\frac{d}{\varepsilon_r(2\lambda + d)}}~,\]
where $L$ is the junction width, $d$ is the tunnel barrier thickness, $\varepsilon_r$ is the dielectric constant of the barrier, and $n$ is the mode index.  For the 15~{\AA} device, the 1st- and 2nd-order Fiske modes occurred at multiples of 0.27~mV.  This voltage agrees quite well with what Gijsbertson \textit{et al.}\cite{Gijsbertsen1993} and Sugiyama \textit{et al.}\cite{Sugiyama1995} report for similar-sized Nb/Al-AlO$_x$/Nb junctions.  $V_1$ = 0.27~mV gives a Josephson frequency of 130~GHz and a phase velocity, $v_{\textrm{ph}}$ = 0.026$c$.  Letting $L$ = 30 $\mu$m, $d$ = 15~{\AA}, and $\varepsilon_r \approx 10$, the resulting value for $2\lambda$ is approximately 2200~{\AA}, in agreement previous reports.\cite{Broom1976,Cucolo1983}

\begin{figure}
  \includegraphics[width=3.25in]{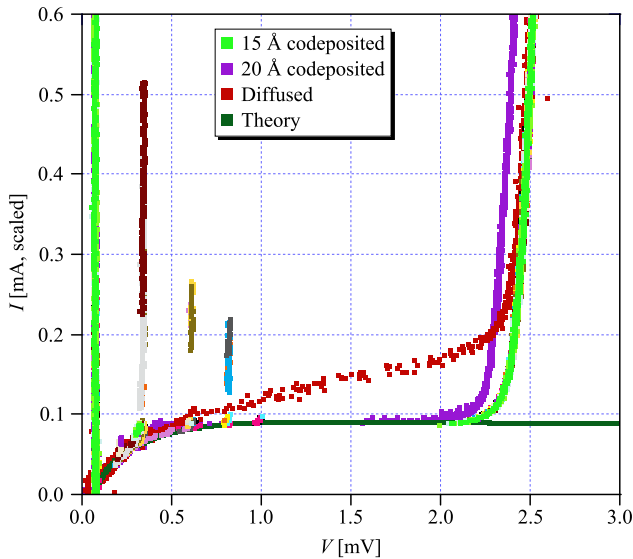}
  \caption{(Color online) $I$-$V$ characteristics of three different $30\times30$ $\mu$m$^2$ junctions expanded to show the subgap current in greater detail: one with 15~{\AA} of codeposited AlO$_x$, one with 20~{\AA} of codeposited AlO$_x$, and one with a diffused barrier.  For comparison, the theoretical prediction is also shown.  Current values have been scaled according to the devices' normal-state resistances.}\label{subgap}
\end{figure}

Junctions with codeposited barriers also show subgap currents that agree with the theoretical prediction of an ideal tunnel junction.  Due to the thermal excitation of quasiparticles at finite temperatures, the subgap current is nonzero, and for the special case where  $\Delta_1$ = $\Delta_2$ = $\Delta$ and $k_BT \ll \Delta$, the subgap current, $I_{sg}$, can be well-approximated by:\cite{VanDuzer1998}
\begin{equation}
I_{sg} = 2G_n \left(V + \frac{\Delta}{e}\right) e^{-\Delta/k_BT} \sqrt{\frac{2\Delta}{eV + 2\Delta}} \sinh(\xi) K_0(\xi)~,\label{Isg}
\end{equation}
where $G_n$ is the normal-state conductance, $K_0$ is the zeroth-order modified Bessel function, and $\xi$ = $eV/2k_BT$.  For a device with 15~{\AA} of codeposited AlO$_x$ both the subgap region of the $I$-$V$ curve and the theoretical prediction based on Eqn. \ref{Isg} using the measured value of $G_n$ are shown in Fig. \ref{subgap}.  Also shown are scaled $I$-$V$ curves for a tunnel junction with 20~{\AA} of codeposited AlO$_x$, and one with a barrier formed by diffusive oxidation of the Al (111) surface.  Both of these curves are scaled according to their respective normal-state resistances.

For the devices with codeposited AlO$_x$ barriers the agreement with theory was somewhat surprising, considering that the condition $\Delta_1$ = $\Delta_2$ was not obviously satisfied due to the 200~{\AA} of Al metal on one side of the junction.  However, the Al layer was epitaxial, having a long mean free path, and the interface with the Nb electrode below was free of chemical impurities and structural defects.  For this reason, the Nb/Al base layer likely acted as a homogeneous superconductor, essentially in the Cooper limit, with an electronic structure dominated at 4.2~K by the Nb film.\cite{deGennes1964}  Indeed, the $I$-$V$ characteristics (Fig. \ref{althickvar}) indicated that a 200~{\AA} Al layer may be thin enough for the measured gap to asymptotically approach the bare Nb value.  In addition, the quantitative agreement with theory suggested that the measured subgap characteristics of these devices were determined by thermal excitation of quasiparticles and not by defects in the trilayer materials.  There was no apparent transport mode that was not directly attributable to ideal superconductivity in the electrodes and lossless tunneling through the barrier at finite bias.  The same could not be said of the tunnel junction having a diffused oxide barrier (Fig. \ref{subgap}, red curve), where a shunt conductance appeared in parallel with the expected subgap current.  This shunt conductance could be attributed to a number of possible factors, some not related to the intrinsic quality of the trilayer materials at all (eg., device fabrication).  However, the fact that none of our Josephson junctions with diffused barriers showed the theoretically predicted subgap current -- while those with grown oxides did -- led to the conclusion that this extra shunt conductance was due to defects in the diffused oxide.

In summary, we have fabricated tunnel junctions from trilayers with barriers formed by different methods.  The base electrodes in all cases were epitaxial Nb/Al bilayers grown on A-plane sapphire.  The tunnel barriers were formed either by diffusive oxidation of the Al surface, or by codepositing Al and O$_2$ on the epitaxial Al layer.  Applied magnetic fields were shown to have the expected effect on $I_c$, and the observation of Fiske modes was an indication of particularly uniform barriers.  Finally, we found that the subgap currents in devices with codeposited barriers agreed well with theory, while those with diffused oxides showed a parallel shunt conductance.  This suggests that codeposition yields a more uniform and higher quality oxide barrier which may better serve the requirements of quantum circuits.  In particular, codeposition may be key to minimizing the density of oxygen vacancies in the barrier and reducing the number of two-level systems observed in superconducting quantum bits.  This has already been demonstrated for the case of crystalline tunnel barriers.\cite{Oh2006a}  Based on our findings, we speculate that codeposited amorphous tunnel barriers may be equally advantageous.

This research was partially funded by the Office of the Director of National Intelligence (ODNI), Intelligence Advanced Research Projects Activity (IARPA), through the Army Research Office.  All statements of fact, opinion or conclusions contained herein are those of the authors and should not be construed as representing the official views or policies of IARPA, the ODNI, or the U.S. Government.  It was also partially funded by the National Science Foundation through Grant No. EIA 01-21568.  AFM and XRD analyses were carried out in the Center for Microanalysis of Materials, University of Illinois at Urbana-Champaign, which is partially supported by the U.S. Department of Energy under Grant No. DEFG02-91ER45439.


\begin{thebibliography}{10}%
\makeatletter
\providecommand \@ifxundefined [1]{%
 \ifx #1\undefined \expandafter \@firstoftwo
 \else \expandafter \@secondoftwo
\fi
}%
\providecommand \@ifnum [1]{%
 \ifnum #1\expandafter \@firstoftwo
 \else \expandafter \@secondoftwo
\fi
}%
\providecommand \enquote [1]{``#1''}%
\providecommand \bibnamefont  [1]{#1}%
\providecommand \bibfnamefont [1]{#1}%
\providecommand \citenamefont [1]{#1}%
\providecommand\href[0]{\@sanitize\@href}%
\providecommand\@href[1]{\endgroup\@@startlink{#1}\endgroup\@@href}%
\providecommand\@@href[1]{#1\@@endlink}%
\providecommand \@sanitize [0]{\begingroup\catcode`\&12\catcode`\#12\relax}%
\@ifxundefined \pdfoutput {\@firstoftwo}{%
 \@ifnum{\z@=\pdfoutput}{\@firstoftwo}{\@secondoftwo}%
}{%
 \providecommand\@@startlink[1]{\leavevmode}%
 \providecommand\@@endlink[0]{}%
}{%
 \providecommand\@@startlink[1]{%
  \leavevmode
  \pdfstartlink
   attr{/Border[0 0 1 ]/H/I/C[0 1 1]}%
   user{/Subtype/Link/A<</Type/Action/S/URI/URI(#1)>>}%
  \relax
 }%
 \providecommand\@@endlink[0]{\pdfendlink}%
}%
\providecommand \url  [0]{\begingroup\@sanitize \@url }%
\providecommand \@url [1]{\endgroup\@href {#1}{\urlprefix}}%
\providecommand \urlprefix [0]{URL }%
\providecommand \Eprint[0]{\href }%
\@ifxundefined \urlstyle {%
  \providecommand \doi [1]{doi:\discretionary{}{}{}#1}%
}{%
  \providecommand \doi [0]{doi:\discretionary{}{}{}\begingroup
  \urlstyle{rm}\Url }%
}%
\providecommand \doibase [0]{http://dx.doi.org/}%
\providecommand \Doi[1]{\href{\doibase#1}}%
\providecommand \selectlanguage [0]{\@gobble}%
\providecommand \bibinfo [0]{\@secondoftwo}%
\providecommand \bibfield [0]{\@secondoftwo}%
\providecommand \translation [1]{[#1]}%
\providecommand \BibitemOpen[0]{}%
\providecommand \bibitemStop [0]{}%
\providecommand \bibitemNoStop [0]{.\EOS\space}%
\providecommand \EOS [0]{\spacefactor3000\relax}%
\providecommand \BibitemShut [1]{\csname bibitem#1\endcsname}%
\bibitem{Josephson1962}%
  \BibitemOpen
  \bibfield{author}{%
  \bibinfo {author} {\bibfnamefont{B.~D.}\ \bibnamefont{Josephson}},\ }%
  \bibfield{journal}{%
  \bibinfo {journal} {Phys. Lett.}\ }%
  \textbf{\bibinfo {volume} {1}},\ \bibinfo {pages} {251} (\bibinfo {year}
  {1962})\BibitemShut{NoStop}%
\bibitem{Likharev1979}%
  \BibitemOpen
  \bibfield{author}{%
  \bibinfo {author} {\bibfnamefont{K.~K.}\ \bibnamefont{Likharev}},\ }%
  \bibfield{journal}{%
  \bibinfo {journal} {Rev. Mod. Phys.}\ }%
  \textbf{\bibinfo {volume} {51}},\ \bibinfo {pages} {101} (\bibinfo {year}
  {1979})\BibitemShut{NoStop}%
\bibitem{Gurvitch1983}%
  \BibitemOpen
  \bibfield{author}{%
  \bibinfo {author} {\bibfnamefont{M.}~\bibnamefont{Gurvitch}}, \bibinfo
  {author} {\bibfnamefont{M.~A.}\ \bibnamefont{Washington}},\ and\ \bibinfo
  {author} {\bibfnamefont{H.~A.}\ \bibnamefont{Huggins}},\ }%
  \bibfield{journal}{%
  \bibinfo {journal} {Appl. Phys. Lett.}\ }%
  \textbf{\bibinfo {volume} {42}},\ \bibinfo {pages} {472} (\bibinfo {year}
  {1983})\BibitemShut{NoStop}%
\bibitem{Simmonds2004}%
  \BibitemOpen
  \bibfield{author}{%
  \bibinfo {author} {\bibfnamefont{R.~W.}\ \bibnamefont{Simmonds}}, \bibinfo
  {author} {\bibfnamefont{K.~M.}\ \bibnamefont{Lang}}, \bibinfo {author}
  {\bibfnamefont{D.~A.}\ \bibnamefont{Hite}}, \bibinfo {author}
  {\bibfnamefont{S.}~\bibnamefont{Nam}}, \bibinfo {author}
  {\bibfnamefont{D.~P.}\ \bibnamefont{Pappas}},\ and\ \bibinfo {author}
  {\bibfnamefont{J.~M.}\ \bibnamefont{Martinis}},\ }%
  \bibfield{journal}{%
  \bibinfo {journal} {Phys. Rev. Lett.}\ }%
  \textbf{\bibinfo {volume} {93}},\ \bibinfo {pages} {077003} (\bibinfo {year}
  {2004})\BibitemShut{NoStop}%
\bibitem{Tan2005}%
  \BibitemOpen
  \bibfield{author}{%
  \bibinfo {author} {\bibfnamefont{E.}~\bibnamefont{Tan}}, \bibinfo {author}
  {\bibfnamefont{P.~G.}\ \bibnamefont{Mather}}, \bibinfo {author}
  {\bibfnamefont{A.~C.}\ \bibnamefont{Perrella}}, \bibinfo {author}
  {\bibfnamefont{J.~C.}\ \bibnamefont{Read}},\ and\ \bibinfo {author}
  {\bibfnamefont{R.~A.}\ \bibnamefont{Buhrman}},\ }%
  \bibfield{journal}{%
  \bibinfo {journal} {Phys. Rev. B}\ }%
  \textbf{\bibinfo {volume} {71}},\ \bibinfo {pages} {161401(R)} (\bibinfo
  {year} {2005})\BibitemShut{NoStop}%
\bibitem{Mather2005}%
  \BibitemOpen
  \bibfield{author}{%
  \bibinfo {author} {\bibfnamefont{P.~G.}\ \bibnamefont{Mather}}, \bibinfo
  {author} {\bibfnamefont{A.~C.}\ \bibnamefont{Perrella}}, \bibinfo {author}
  {\bibfnamefont{E.}~\bibnamefont{Tan}}, \bibinfo {author}
  {\bibfnamefont{J.~C.}\ \bibnamefont{Read}},\ and\ \bibinfo {author}
  {\bibfnamefont{R.~A.}\ \bibnamefont{Buhrman}},\ }%
  \bibfield{journal}{%
  \bibinfo {journal} {Appl. Phys. Lett.}\ }%
  \textbf{\bibinfo {volume} {86}},\ \bibinfo {pages} {242504} (\bibinfo {year}
  {2005})\BibitemShut{NoStop}%
\bibitem{VanDuzer1998}%
  \BibitemOpen
  \bibfield{author}{%
  \bibinfo {author} {\bibfnamefont{T.~V.}\ \bibnamefont{Duzer}}\ and\ \bibinfo
  {author} {\bibfnamefont{C.~W.}\ \bibnamefont{Turner}},\ }%
  \emph{\bibinfo {title} {Principles of Superconductive Devices and Circuits}}\
  (\bibinfo {publisher} {Prentice Hall PTR},\ \bibinfo {address} {Upper Saddle
  River, NJ},\ \bibinfo {year} {1998})\BibitemShut{NoStop}%
\bibitem{Welander2007}%
  \BibitemOpen
  \bibfield{author}{%
  \bibinfo {author} {\bibfnamefont{P.~B.}\ \bibnamefont{Welander}}\ and\
  \bibinfo {author} {\bibfnamefont{J.~N.}\ \bibnamefont{Eckstein}},\ }%
  \bibfield{journal}{%
  \bibinfo {journal} {Appl. Phys. Lett.}\ }%
  \textbf{\bibinfo {volume} {90}},\ \bibinfo {pages} {243510} (\bibinfo {year}
  {2007})\BibitemShut{NoStop}%
\bibitem{WelanderThesis}%
  \BibitemOpen
  \bibfield{author}{%
  \bibinfo {author} {\bibfnamefont{P.~B.}\ \bibnamefont{Welander}},\ }%
  Ph.D. thesis,\ \bibinfo {school} {Univ. of Illinois at Urbana-Champaign}
  (\bibinfo {year} {2007})\BibitemShut{NoStop}%
\bibitem{Bruce1978}%
  \BibitemOpen
  \bibfield{author}{%
  \bibinfo {author} {\bibfnamefont{L.~A.}\ \bibnamefont{Bruce}}\ and\ \bibinfo
  {author} {\bibfnamefont{H.}~\bibnamefont{Jaeger}},\ }%
  \bibfield{journal}{%
  \bibinfo {journal} {Phil. Mag. A}\ }%
  \textbf{\bibinfo {volume} {38}},\ \bibinfo {pages} {223} (\bibinfo {year}
  {1978})\BibitemShut{NoStop}%
\bibitem{Tao1984}%
  \BibitemOpen
  \bibfield{author}{%
  \bibinfo {author} {\bibfnamefont{H.~J.}\ \bibnamefont{Tao}}, \bibinfo
  {author} {\bibfnamefont{E.~D.}\ \bibnamefont{Gibson}}, \bibinfo {author}
  {\bibfnamefont{J.~D.}\ \bibnamefont{Verhoeven}},\ and\ \bibinfo {author}
  {\bibfnamefont{E.~L.}\ \bibnamefont{Wolf}},\ }%
  \bibfield{journal}{%
  \bibinfo {journal} {Phil. Mag. B}\ }%
  \textbf{\bibinfo {volume} {50}},\ \bibinfo {pages} {L55} (\bibinfo {year}
  {1984})\BibitemShut{NoStop}%
\bibitem{Yamamoto2002}%
  \BibitemOpen
  \bibfield{author}{%
  \bibinfo {author} {\bibfnamefont{S.}~\bibnamefont{Yamamoto}}\ and\ \bibinfo
  {author} {\bibfnamefont{H.}~\bibnamefont{Naramoto}},\ }%
  \bibfield{journal}{%
  \bibinfo {journal} {Nucl. Instr. and Meth. B}\ }%
  \textbf{\bibinfo {volume} {190}},\ \bibinfo {pages} {657} (\bibinfo {year}
  {2002})\BibitemShut{NoStop}%
\bibitem{Braginski1986}%
  \BibitemOpen
  \bibfield{author}{%
  \bibinfo {author} {\bibfnamefont{A.~I.}\ \bibnamefont{Braginski}}, \bibinfo
  {author} {\bibfnamefont{J.}~\bibnamefont{Talvacchio}}, \bibinfo {author}
  {\bibfnamefont{M.~A.}\ \bibnamefont{Janocko}},\ and\ \bibinfo {author}
  {\bibfnamefont{J.~R.}\ \bibnamefont{Gavaler}},\ }%
  \bibfield{journal}{%
  \bibinfo {journal} {J. Appl. Phys.}\ }%
  \textbf{\bibinfo {volume} {60}},\ \bibinfo {pages} {2058} (\bibinfo {year}
  {1986})\BibitemShut{NoStop}%
\bibitem{Fiske1964}%
  \BibitemOpen
  \bibfield{author}{%
  \bibinfo {author} {\bibfnamefont{M.~D.}\ \bibnamefont{Fiske}},\ }%
  \bibfield{journal}{%
  \bibinfo {journal} {Rev. Mod. Phys.}\ }%
  \textbf{\bibinfo {volume} {36}},\ \bibinfo {pages} {221} (\bibinfo {year}
  {1964})\BibitemShut{NoStop}%
\bibitem{Eck1964}%
  \BibitemOpen
  \bibfield{author}{%
  \bibinfo {author} {\bibfnamefont{R.~E.}\ \bibnamefont{Eck}}, \bibinfo
  {author} {\bibfnamefont{D.~J.}\ \bibnamefont{Scalapino}},\ and\ \bibinfo
  {author} {\bibfnamefont{B.~N.}\ \bibnamefont{Taylor}},\ }%
  \bibfield{journal}{%
  \bibinfo {journal} {Phys. Rev. Lett.}\ }%
  \textbf{\bibinfo {volume} {13}},\ \bibinfo {pages} {15} (\bibinfo {year}
  {1964})\BibitemShut{NoStop}%
\bibitem{Kulik1965}%
  \BibitemOpen
  \bibfield{author}{%
  \bibinfo {author} {\bibfnamefont{I.~O.}\ \bibnamefont{Kulik}},\ }%
  \bibfield{journal}{%
  \bibinfo {journal} {Pis'ma Zh. \'{E}ksp. Teor. Fiz.}\ }%
  \textbf{\bibinfo {volume} {2}},\ \bibinfo {pages} {134} (\bibinfo {year}
  {1965}); [JETP Lett. \textbf{2}, 84 (1965)]\BibitemShut{NoStop}%
\bibitem{Gijsbertsen1993}%
  \BibitemOpen
  \bibfield{author}{%
  \bibinfo {author} {\bibfnamefont{J.~G.}\ \bibnamefont{Gijsbertsen}}, \bibinfo
  {author} {\bibfnamefont{E.~P.}\ \bibnamefont{Houwman}}, \bibinfo {author}
  {\bibfnamefont{J.}~\bibnamefont{Flokstra}}, \bibinfo {author}
  {\bibfnamefont{H.}~\bibnamefont{Rogalla}}, \bibinfo {author}
  {\bibfnamefont{J.~B.}\ \bibnamefont{le~Grand}},\ and\ \bibinfo {author}
  {\bibfnamefont{P.~A.~J.}\ \bibnamefont{de~Korte}},\ }%
  \bibfield{journal}{%
  \bibinfo {journal} {IEEE Trans. Appl. Supercond.}\ }%
  \textbf{\bibinfo {volume} {3}},\ \bibinfo {pages} {2100} (\bibinfo {year}
  {1993})\BibitemShut{NoStop}%
\bibitem{Sugiyama1995}%
  \BibitemOpen
  \bibfield{author}{%
  \bibinfo {author} {\bibfnamefont{H.}~\bibnamefont{Sugiyama}}, \bibinfo
  {author} {\bibfnamefont{A.}~\bibnamefont{Fujimaki}},\ and\ \bibinfo {author}
  {\bibfnamefont{H.}~\bibnamefont{Hayakawa}},\ }%
  \bibfield{journal}{%
  \bibinfo {journal} {IEEE Trans. Appl. Supercond.}\ }%
  \textbf{\bibinfo {volume} {5}},\ \bibinfo {pages} {2739} (\bibinfo {year}
  {1995})\BibitemShut{NoStop}%
\bibitem{Broom1976}%
  \BibitemOpen
  \bibfield{author}{%
  \bibinfo {author} {\bibfnamefont{R.~F.}\ \bibnamefont{Broom}},\ }%
  \bibfield{journal}{%
  \bibinfo {journal} {J. Appl. Phys.}\ }%
  \textbf{\bibinfo {volume} {47}},\ \bibinfo {pages} {5432} (\bibinfo {year}
  {1976})\BibitemShut{NoStop}%
\bibitem{Cucolo1983}%
  \BibitemOpen
  \bibfield{author}{%
  \bibinfo {author} {\bibfnamefont{A.~M.}\ \bibnamefont{Cucolo}}, \bibinfo
  {author} {\bibfnamefont{S.}~\bibnamefont{Pace}}, \bibinfo {author}
  {\bibfnamefont{R.}~\bibnamefont{Vaglio}}, \bibinfo {author}
  {\bibfnamefont{A.~D.}\ \bibnamefont{Chiara}}, \bibinfo {author}
  {\bibfnamefont{G.}~\bibnamefont{Peluso}},\ and\ \bibinfo {author}
  {\bibfnamefont{M.}~\bibnamefont{Russo}},\ }%
  \bibfield{journal}{%
  \bibinfo {journal} {J. Low Temp. Phys.}\ }%
  \textbf{\bibinfo {volume} {50}},\ \bibinfo {pages} {301} (\bibinfo {year}
  {1983})\BibitemShut{NoStop}%
\bibitem{deGennes1964}%
  \BibitemOpen
  \bibfield{author}{%
  \bibinfo {author} {\bibfnamefont{P.~G.}\ \bibnamefont{de Gennes}},\ }%
  \bibfield{journal}{%
  \bibinfo {journal} {Rev. Mod. Phys.}\ }%
  \textbf{\bibinfo {volume} {36}},\ \bibinfo {pages} {225} (\bibinfo {year}
  {1964})\BibitemShut{NoStop}%
\bibitem{Oh2006a}%
  \BibitemOpen
  \bibfield{author}{%
  \bibinfo {author} {\bibfnamefont{S.}~\bibnamefont{Oh}}, \bibinfo {author}
  {\bibfnamefont{K.}~\bibnamefont{Cicak}}, \bibinfo {author}
  {\bibfnamefont{J.~S.}\ \bibnamefont{Kline}}, \bibinfo {author}
  {\bibfnamefont{M.~A.}\ \bibnamefont{Sillanp\"{a}\"{a}}}, \bibinfo {author}
  {\bibfnamefont{K.~D.}\ \bibnamefont{Osborn}}, \bibinfo {author}
  {\bibfnamefont{J.~D.}\ \bibnamefont{Whittaker}}, \bibinfo {author}
  {\bibfnamefont{R.~W.}\ \bibnamefont{Simmonds}},\ and\ \bibinfo {author}
  {\bibfnamefont{D.~P.}\ \bibnamefont{Pappas}},\ }%
  \bibfield{journal}{%
  \bibinfo {journal} {Phys. Rev. B}\ }%
  \textbf{\bibinfo {volume} {74}},\ \bibinfo {pages} {100502} (\bibinfo {year}
  {2006})\BibitemShut{NoStop}%
\end{thebibliography}

%

\end{document}